# Maxwell's macroscopic equations, the energy-momentum postulates, and the Lorentz law of force

Masud Mansuripur[†] and Armis R. Zakharian[‡]

[†]College of Optical Sciences, The University of Arizona, Tucson, Arizona 85721
[‡]Corning Incorporated, Science and Technology Division, Corning, New York 14831

**Abstract**. We argue that the classical theory of electromagnetism is based on Maxwell's macroscopic equations, an energy postulate, a momentum postulate, and a generalized form of the Lorentz law of force. These seven postulates constitute the foundation of a complete and consistent theory, thus eliminating the need for actual (i.e., physical) models of polarization ***P*** and magnetization ***M***, these being the distinguishing features of Maxwell's macroscopic equations. In the proposed formulation, ***P***(***r***, *t*) and ***M***(***r***, *t*) are arbitrary functions of space and time, their physical properties being embedded in the seven postulates of the theory. The postulates are self-consistent, comply with the requirements of the special theory of relativity, and satisfy the laws of conservation of energy, linear momentum, and angular momentum. One advantage of the proposed formulation is that it side-steps the long-standing Abraham-Minkowski controversy surrounding the electromagnetic momentum inside a material medium by simply "assigning" the Abraham momentum density ***E***(***r***,*t*)×***H***(***r***,*t*)/$c^2$ to the electromagnetic field. This well-defined momentum is thus taken to be universal as it does not depend on whether the field is propagating or evanescent, and whether or not the host medium is homogeneous, transparent, isotropic, dispersive, magnetic, linear, etc. In other words, the local and instantaneous momentum density is uniquely and unambiguously specified at each and every point of the material system in terms of the ***E*** and ***H*** fields residing at that point. Any variation with time of the *total* electromagnetic momentum of a closed system results in a force exerted on the material media within the system in accordance with the generalized Lorentz law.

**1. Introduction**. Maxwell's macroscopic equations are mathematically precise, self-consistent, and fully compatible with the special theory of relativity; however, they require additional postulates to make them complete as well as consistent with the laws of conservation of energy, momentum, and angular momentum. In addition to the densities of free charge and free current, $\rho_{\text{free}}$ and ***J***$_{\text{free}}$, which are the sources of ***E*** and ***H*** fields in the *microscopic* equations, Maxwell's macroscopic equations incorporate the polarization density ***P*** and the magnetization density ***M*** as additional sources of the electromagnetic field [1,2]. We emphasize in this paper that there is no need for explicit physical models or interpretations of ***P***(***r***, *t*) and ***M***(***r***, *t*); rather, these should be treated as well-behaved functions of space and time that obey certain restrictions imposed upon them by special relativity and by the enunciated postulates of the theory of electromagnetism. In this view of the classical theory, it is the *macroscopic* Maxwell equations that are fundamental, reducing to the simpler *microscopic* equations when ***P*** and ***M*** vanish. Our discussions of electromagnetic energy, momentum, force and torque in the following sections hint at the strong possibility that the properties of ***P*** and ***M*** may not, after all, be deducible from those of bound charges and currents (as is done in conventional treatments by invoking standard models of dielectric polarization and magnetization). Thus the claim that Maxwell's macroscopic equations, being broader in scope, are also more fundamental than his microscopic equations may not, in our view, be a matter of convenience but, rather, a deep-rooted statement concerning the physics of electromagnetism.

As is well-known, the macroscopic equations in conjunction with the definitions ***D*** = $\varepsilon_o$***E*** + ***P*** and ***B*** = $\mu_o$***H*** + ***M*** relate the four fields ***E***, ***D***, ***H***, ***B*** to their sources $\rho_{\text{free}}$, ***J***$_{\text{free}}$, ***P*** and ***M*** [3,4]; here $\varepsilon_o$ and $\mu_o$ are the permittivity and permeability of free space. The system of units used throughout the paper is MKSA.

In the absence of specific models or assumptions regarding the physical nature of ***P*** and ***M***, it becomes necessary to postulate the relation between the fields and their energy content. It turns out that the only required postulate of the theory concerning energy is the statement relating the time-rate-of-change of energy density to the local fields and their time-derivatives; see Eq.(11). With this postulate in hand, it is readily shown in the most general case that the rate of flow of electromagnetic energy (per unit area per unit time) is the Poynting vector ***S***(***r***,*t*)= ***E***(***r***,*t*)×***H***(***r***,*t*). The energy postulate, in conjunction with Maxwell's macroscopic equations, is fully consistent with the law of conservation of energy. It can be shown that any energy entering a closed volume is either stored in the fields or consumed in the interaction between the fields and the sources located within the volume. Similarly, any energy exiting a closed volume is either released from the fields or generated as a result of interactions between the fields and the sources internal to the volume.

The momentum density ***p***$_{\text{EM}}$(***r***, *t*) of the electromagnetic field is another fundamental entity that needs explicit postulation. While there exist physical arguments for deriving from first principles the momentum of a propagating field in vacuum [2], there remains a long-running controversy as to the nature of the field's momentum inside material media – the well-known Abraham-Minkowski controversy [5-7]. We believe that under general circumstances the field's momentum cannot be derived from first principles, especially in the absence of physical models that pin down the essence of $P$ and $M$. It thus becomes necessary to resort to postulating the field's momentum density; our postulated expression is $p_{EM}(r, t) = S(r,t)/c^2$; see Eqs. (13). This expression, known to be valid for propagating waves in free space, also appears to hold for static fields [2], for evanescent fields, and for fields within media generally specified in terms of $\rho_{free}$, $J_{free}$, $P$ and $M$, whether or not these host media are transparent, partially absorptive, dispersive, birefringent, magnetic, non-linear, etc. [8-19]. In other words, the momentum density of the electromagnetic field under *all* circumstances is simply the local Poynting vector normalized by the square of the speed of light in vacuum.

So long as an electromagnetic field distribution remains entirely in free space (or entirely within a homogeneous, transparent medium), its total momentum remains constant. However, once the field encounters a change in the environment, or begins to get scattered and/or absorbed, its momentum begins to vary with time. As an example, consider a finite-length, finite-diameter pulse of light propagating in free space. The shape of the pulse will change as it propagates, but its total momentum (i.e., integrated momentum density) remains constant in time. Now, if the pulse arrives at a massive, perfectly reflecting, flat mirror, say, at normal incidence, it will be reflected without any loss of energy. At first, there will be some overlap between the incident and reflected beams, but eventually the propagation direction of the incident pulse will be fully reversed. While the pulse interacts with the mirror, its electromagnetic momentum varies with time, eventually settling at the opposite of its initial value. This change of momentum is accompanied by a temporary force exerted on the mirror, resulting in the transfer of twice the field's initial momentum to the mirror (in the form of mechanical momentum); the momentum transfer is thus mediated by the exerted force. In order for the momentum of the entire system to be conserved, it is essential that the instantaneous force experienced by the mirror be precisely equal to the time-rate-of-change of the field's momentum at each instant of time.

The above example may be generalized by replacing the mirror with an arbitrary medium (not necessarily a reflector), whose electromagnetic properties are completely specified in terms of $\rho_{free}$, $J_{free}$, $P$ and $M$. Once again, under any and all circumstances, the force of the electromagnetic field on the material medium must be exactly equal to the time-rate-of-change of the field's total momentum, lest the momentum conservation law is violated. (Similarly, the torque exerted by the electromagnetic field on the medium must be identical with the time-rate-of-change of the field's total angular momentum, or else, conservation of angular momentum will be in jeopardy.) Clearly, the postulate expressing the field's momentum in terms of its Poynting vector is not arbitrary; rather, within the region occupied by the field, an intimate connection must exist between the field's momentum and the electromagnetic force exerted on the material medium [18, 19].

The stage is now set for introducing the last postulate of the classical theory of electromagnetism. Historically, this last postulate has been called the Lorentz law of force and expressed in the form of $F = q(E + V \times B)$, where a point charge $q$ moving with velocity $V$ experiences the force $F$ from the local $E$ and $B$ fields [1-3]. While this expression can be readily written in terms of the free charge and current densities ($\rho_{free}$, $J_{free}$), its extension to cover media that contain $P$ and $M$ is problematic. Traditionally, models have been devised in the form of dense aggregates of atomic electric dipoles (for $P$), and dense aggregates of infinitesimal electric current loops (for $M$) – the so-called Amperian model. Subsequently, the force law has been extended to media that exhibit polarization $P$ and/or magnetization $M$ [3, 20-24]. These models are highly complex, require heroic efforts to account for self-interactions, and tend to ignore the quantum nature of atomic polarization and magnetization. We believe a better approach to the force law is simply to postulate a generalization of the Lorentz expression that explicitly includes the contributions of $\rho_{free}$, $J_{free}$, $P$ and $M$. Such a generalization, of course, cannot be made arbitrarily; it must conform with the conservation laws, with the special theory of relativity, and with the aforesaid postulates concerning the densities of electromagnetic energy and momentum. As it turns out, there exist not one but (as far as we know) two possible formulations of the generalized force law that satisfy the above requirements [3, 16]. One such expression for force density (along with its companion for torque density) is given by Eqs. (14), the other by Eqs. (15). It can be shown that the total force (and total torque) on a given object in the presence of an electromagnetic field is the same, no matter which expression is used [3, 25-28]. However, the *distribution* of force (and torque) throughout the body of the object will be different for the two expressions.

In the following sections we present a brief summary of the seven postulates that form the foundations of the classical theory of electromagnetism. In addition to the four macroscopic equations of Maxwell, Eqs. (1), these include an expression for the time-rate-of-change of energy density, Eq. (11), the postulate of electromagnetic momentum density, Eqs. (13), and the generalized Lorentz law in the form of Eqs. (14) or Eqs. (15). No assumptions will be necessary to ascertain the nature of $P$ and $M$ beyond the conventional definitions, Eqs. (2), and the Lorentz rules for transforming $P$ and $M$ between inertial frames, Eqs. (6). The conditions under which electromagnetic momentum and energy constitute a relativistic 4-vector are explored in Section 11. Finally, in Section 12 we present the results of numerical simulations that illustrate the intimate connection between the time-rate-of-change of electromagnetic momentum in a closed system and the total force exerted on the material media within that system.

**2. Maxwell's macroscopic equations**. In the MKSA system of units, Maxwell's macroscopic equations are

$$\nabla \cdot \boldsymbol{D} = \rho_{\text{free}}, \tag{1a}$$

$$\nabla \times \boldsymbol{H} = \boldsymbol{J}_{\text{free}} + \partial \boldsymbol{D}/\partial t, \tag{1b}$$

$$\nabla \times \boldsymbol{E} = -\partial \boldsymbol{B}/\partial t, \tag{1c}$$

$$\nabla \cdot \boldsymbol{B} = 0. \tag{1d}$$

In these equations, electric displacement $\boldsymbol{D}$ and magnetic induction $\boldsymbol{B}$ are related to the polarization density $\boldsymbol{P}$ and magnetization density $\boldsymbol{M}$ via the identities

$$\boldsymbol{D} = \varepsilon_0 \boldsymbol{E} + \boldsymbol{P}, \tag{2a}$$

$$\boldsymbol{B} = \mu_0 \boldsymbol{H} + \boldsymbol{M}. \tag{2b}$$

In general, $\rho_{\text{free}}$, $\boldsymbol{J}_{\text{free}}$, $\boldsymbol{P}$, $\boldsymbol{M}$, $\boldsymbol{E}$, $\boldsymbol{H}$, $\boldsymbol{D}$, and $\boldsymbol{B}$ appearing in the above equations are functions of space and time $(\boldsymbol{r}, t)$ specified in an inertial frame of reference. The free charge and current densities, of course, satisfy the continuity equation, $\nabla \cdot \boldsymbol{J}_{\text{free}} + \partial \rho_{\text{free}}/\partial t = 0$, and together they form a 4-vector $(\boldsymbol{J}_{\text{free}}, c\rho_{\text{free}})$ that transforms between inertial frames in accordance with the Lorentz transformation rules of special relativity [1-3].

Note that Eqs.(2) are neither new assumptions nor independent postulates; they simply *define* the $\boldsymbol{D}$ and $\boldsymbol{B}$ fields, which made their first appearance in Eqs.(1). A good way to approach Maxwell's equations then is to recognize that, while the so-called microscopic theory limits the sources of the $\boldsymbol{E}, \boldsymbol{D}, \boldsymbol{H}, \boldsymbol{B}$ fields to $\rho_{\text{free}}$ and $\boldsymbol{J}_{\text{free}}$, the macroscopic theory is enormously enriched by the addition of $\boldsymbol{P}$ and $\boldsymbol{M}$ as two essentially independent sources of the electromagnetic field.

**3. Bound electric charge-density and current-density arising from $\boldsymbol{P}$ and $\boldsymbol{M}$**. One can *define* bound electric charge and current densities, $\rho_{\text{e\_bound}} = -\nabla \cdot \boldsymbol{P}(\boldsymbol{r}, t)$ and $\boldsymbol{J}_{\text{e\_bound}} = \partial \boldsymbol{P}(\boldsymbol{r}, t)/\partial t$, arising from the polarization $\boldsymbol{P}$, as well as an effective electric current density $\boldsymbol{J}_{\text{e\_mag}} = \mu_0^{-1} \nabla \times \boldsymbol{M}$, that gives rise to the magnetization $\boldsymbol{M}$. Subsequently, the macroscopic equations (1) may be written in the following equivalent way:

$$\varepsilon_0 \nabla \cdot \boldsymbol{E} = \rho_{\text{free}} + \rho_{\text{e\_bound}}, \tag{3a}$$

$$\nabla \times \boldsymbol{B} = \mu_0 (\boldsymbol{J}_{\text{free}} + \boldsymbol{J}_{\text{e\_bound}} + \boldsymbol{J}_{\text{e\_mag}}) + \mu_0 \varepsilon_0 \partial \boldsymbol{E}/\partial t, \tag{3b}$$

$$\nabla \times \boldsymbol{E} = -\partial \boldsymbol{B}/\partial t, \tag{3c}$$

$$\nabla \cdot \boldsymbol{B} = 0. \tag{3d}$$

It is now possible to obtain (in the Lorentz gauge) the scalar and vector potentials $\psi(\boldsymbol{r}, t)$ and $\boldsymbol{A}(\boldsymbol{r}, t)$ as integrals over the total charge and current densities, $\rho(\boldsymbol{r}, t) = \rho_{\text{free}} + \rho_{\text{e\_bound}}$ and $\boldsymbol{J}(\boldsymbol{r}, t) = \boldsymbol{J}_{\text{free}} + \boldsymbol{J}_{\text{e\_bound}} + \boldsymbol{J}_{\text{e\_mag}}$, that is,

$$\psi(\boldsymbol{r}, t) = (4\pi\varepsilon_0)^{-1} \iiint \{\rho[\boldsymbol{r}', t-|\boldsymbol{r}-\boldsymbol{r}'|/c]/|\boldsymbol{r}-\boldsymbol{r}'|\} \mathrm{d}v', \tag{4a}$$

$$\boldsymbol{A}(\boldsymbol{r}, t) = (\mu_0/4\pi) \iiint \{\boldsymbol{J}[\boldsymbol{r}', t-|\boldsymbol{r}-\boldsymbol{r}'|/c]/|\boldsymbol{r}-\boldsymbol{r}'|\} \mathrm{d}v'. \tag{4b}$$

The Lorentz gauge, of course, is the identity relating the scalar and vector potentials, namely,

$$\nabla \cdot \boldsymbol{A}(\boldsymbol{r}, t) + (1/c^2) \partial \psi(\boldsymbol{r}, t)/\partial t = 0. \tag{4c}$$

Once the potentials are found, the $E$- and $B$-fields may be determined as follows:

$$\boldsymbol{E}(\boldsymbol{r}, t) = -\nabla \psi(\boldsymbol{r}, t) - \partial \boldsymbol{A}(\boldsymbol{r}, t)/\partial t, \tag{5a}$$

$$\boldsymbol{B}(\boldsymbol{r}, t) = \nabla \times \boldsymbol{A}(\boldsymbol{r}, t). \tag{5b}$$

The 4-potential $(A_x, A_y, A_z, \psi/c)$, specified in the Lorentz gauge via Eqs.(4), may be transformed from one inertial frame to another in accordance with the Lorentz transformation rules of special relativity [1,2]. A straightforward method of transforming the $E$- and $B$-fields between inertial frames consists of first transforming the 4-potential, followed by deriving the fields from the transformed potentials using Eqs.(5).

**4. Lorentz transformation of $\boldsymbol{P}$ and $\boldsymbol{M}$ between inertial frames**. It is important to recognize that the bound charges and currents that appear in Eqs.(3) must form the 4-vector $(\boldsymbol{J}_{\text{e\_bound}} + \boldsymbol{J}_{\text{e\_mag}}, c\rho_{\text{e\_bound}})$ that obeys not only the continuity equation but also the Lorentz transformation rules. That the continuity equation is satisfied is readily demonstrated from the definitions of $J_{\text{e\_bound}}$, $J_{\text{e\_mag}}$, and $\rho_{\text{e\_bound}}$. The latter constraint, however, dictates the following rules for transforming $P(r,t)$ and $M(r,t)$ in one inertial frame to $P'(r',t')$ and $M'(r',t')$ in another:

$$P'_x = P_x; \qquad P'_y = \gamma(P_y - \varepsilon_o V M_z); \qquad P'_z = \gamma(P_z + \varepsilon_o V M_y). \tag{6a}$$

$$M'_x = M_x; \qquad M'_y = \gamma(M_y + \mu_o V P_z); \qquad M'_z = \gamma(M_z - \mu_o V P_y). \tag{6b}$$

Here the inertial frame specified by $(r,t)$ moves at a constant velocity $V$ along the $x$-axis relative to the frame specified by $(r',t')$. As usual, $\gamma = 1/\sqrt{1-(V/c)^2}$, and the space-time coordinates are transformed as follows:

$$x' = \gamma(x + Vt); \qquad y' = y; \qquad z' = z; \qquad t' = \gamma(t + Vx/c^2). \tag{7}$$

We emphasize once again that, in arriving at the relativistic transformation rules of Eqs.(6), the physical mechanisms responsible for $P$ and $M$ were *not* taken into consideration. Simply stated, the above restrictions are imposed on $P(r,t)$ and $M(r,t)$ by requiring the relativistic invariance of Maxwell's macroscopic equations.

**5. Bound magnetic charge and current densities arising from *P* and *M*.** Maxwell's equations (1) may also be written in terms of bound *magnetic* charge and current densities, $\rho_{\text{m\_bound}} = -\nabla \cdot M(r,t)$ and $J_{\text{m\_bound}} = \partial M(r,t)/\partial t$, as well as an effective magnetic current density $J_{\text{m\_pol}} = -\varepsilon_o^{-1} \nabla \times P$ that may be said to give rise to polarization $P$. The sources in this case are $\rho_{\text{free}}$, $J_{\text{free}}$, $\rho_{\text{m\_bound}}$, and $J_{\text{m\_bound}} + J_{\text{m\_pol}}$, while the fields are $D$ and $H$, as follows:

$$\nabla \cdot D = \rho_{\text{free}}, \tag{8a}$$

$$\nabla \times H = J_{\text{free}} + \partial D/\partial t, \tag{8b}$$

$$\nabla \times D = -\varepsilon_o(J_{\text{m\_bound}} + J_{\text{m\_pol}}) - \mu_o \varepsilon_o \partial H/\partial t, \tag{8c}$$

$$\mu_o \nabla \cdot H = \rho_{\text{m\_bound}}. \tag{8d}$$

In addition to $\rho_{\text{free}}$ and $J_{\text{free}}$, the sources are now specified in the form of magnetic charge and current densities, from which $D(r,t)$ and $H(r,t)$ can be readily determined. The disadvantage of Eqs.(8) over Eqs.(3) is that, because $\nabla \cdot H \neq 0$, a vector potential for the $H$-field can no longer be defined. However, if the fields produced by $\rho_{\text{free}}$ and $J_{\text{free}}$ are treated separately, it will become possible to solve Eqs.(8) for $D$ and $H$ in terms of the magnetic charge and current densities. This is done, in analogy with Eqs.(4), by taking advantage of the fact that $\nabla \cdot D = 0$, then introducing scalar and vector potentials produced by $\rho_{\text{m\_bound}}$ and $(J_{\text{m\_bound}} + J_{\text{m\_pol}})$, respectively.

It is important to recognize that $(J_{\text{m\_bound}} + J_{\text{m\_pol}}, c\rho_{\text{m\_bound}})$ is a 4-vector that obeys the continuity equation as well as the Lorentz transformation rules. The continuity equation is guaranteed by taking the divergence of Eq.(8c), then substituting for $\nabla \cdot H$ from Eq.(8d). Compliance with the Lorentz transformation rules is assured in light of the transformation relations given by Eqs.(6).

**6. The nature of *P* and *M* appearing in Maxwell's macroscopic equations.** The arguments advanced in the preceding sections require no specific knowledge of the physical mechanisms that give rise to $P$ and $M$. All one needs to know is that $P(r,t)$ and $M(r,t)$ are sufficiently well-behaved functions of space and time whose spatial and temporal derivatives may be used to define the effective charge and current densities $\rho_{\text{bound}}$, $J_{\text{bound}}$, etc. Even the presence of finite discontinuities in these functions (e.g., at media boundaries) does not pose serious mathematical obstacles as the discontinuities can be handled through the use of Dirac's delta function. The sole physical constraint on $P$ and $M$ is that they must abide by the transformation rules of Eqs.(6).

In general, the time-dependence of the functions $\rho_{\text{free}}$, $J_{\text{free}}$, $P$, $M$, $E$, $H$, $D$, $B$ can be Fourier transformed into the frequency-domain; for example, the Fourier transform $\mathcal{E}(r,\omega)$ of $E(r,t)$ is given by

$$\mathcal{E}(r,\omega) = \int_{-\infty}^{\infty} E(r,t) \exp(-\mathrm{i}\omega t)\,\mathrm{d}t. \tag{9}$$

In many situations arising in practice, the polarization and magnetization densities $\mathcal{P}(r,\omega)$ and $\mathcal{M}(r,\omega)$ are simply proportional to the local fields $\mathcal{E}(r,\omega)$ and $\mathcal{H}(r,\omega)$, respectively. The proportionality constants are then denoted by $\varepsilon_o \chi_e(\omega)$ and $\mu_o \chi_m(\omega)$, and the frequency-domain $\mathcal{D}$- and $\mathcal{B}$-fields are written

$$\mathcal{D}(r,\omega) = \varepsilon_o(1 + \chi_e)\mathcal{E} = \varepsilon_o \varepsilon(\omega)\mathcal{E}(r,\omega), \tag{10a}$$

$$\mathcal{B}(r,\omega) = \mu_o(1 + \chi_m)\mathcal{H} = \mu_o \mu(\omega)\mathcal{H}(r,\omega). \tag{10b}$$

Homogeneous, linear, isotropic media are thus fully specified by their permittivity $\varepsilon(\omega) = \varepsilon' + \mathrm{i}\varepsilon''$ and permeability $\mu(\omega) = \mu' + \mathrm{i}\mu''$. Any loss of energy in such media will be associated with $\varepsilon''$ and $\mu''$, which, by convention, are ≥ 0. The real parts of $\varepsilon(\omega)$ and $\mu(\omega)$, however, may be positive or negative; in particular, in the case of negative-index media, $\varepsilon' < 0$ and $\mu' < 0$.

In spite of the simplifications afforded by restricting Maxwell's equations to linear media, in what follows we shall avoid such restrictions, thus maintaining the generality of $\boldsymbol{P}(\boldsymbol{r},t)$ and $\boldsymbol{M}(\boldsymbol{r},t)$ as arbitrary functions of space and time, which, nevertheless, abide by the relativistic transformation rules of Eqs.(6). The following discussions concerning energy, momentum, force, and torque are therefore quite general and do *not* depend on any assumptions with regard to homogeneity, isotropy, or linearity as expressed, for example, by Eqs.(10). The functions $\boldsymbol{P}(\boldsymbol{r},t)$ and $\boldsymbol{M}(\boldsymbol{r},t)$ could thus depend on $\boldsymbol{E}(\boldsymbol{r},t)$ and $\boldsymbol{H}(\boldsymbol{r},t)$ in complicated, non-local, non-linear ways, or they may not depend on the fields at all.

**7. Energy of the electromagnetic field**. The field's energy density $\mathcal{E}_{\text{field}}(\boldsymbol{r},t)$ will vary with time when the local $E$-field acts upon the free current density $\boldsymbol{J}_{\text{free}}$, or when the local $D$-field undergoes a change in the presence of $\boldsymbol{E}(\boldsymbol{r},t)$, or when the local $B$-field varies in the presence of $\boldsymbol{H}(\boldsymbol{r},t)$. The complete expression for the time-rate-of-change of the local energy density of the field is

$$\frac{\partial \mathcal{E}_{\text{field}}(\boldsymbol{r},t)}{\partial t} = \boldsymbol{E} \cdot \boldsymbol{J}_{\text{free}} + \boldsymbol{E} \cdot \partial \boldsymbol{D}/\partial t + \boldsymbol{H} \cdot \partial \boldsymbol{B}/\partial t. \tag{11}$$

The similarity of the symbols used to denote the electric field $\boldsymbol{E}(\boldsymbol{r},t)$, its Fourier transform $\mathcal{E}(\boldsymbol{r},\omega)$, and the energy density $\mathcal{E}_{\text{field}}(\boldsymbol{r},t)$, a scalar function of $\boldsymbol{r}$ and $t$, will hopefully not cause confusion. Moreover, it should be emphasized that Eq.(11) does *not* imply that the local, instantaneous energy density $\mathcal{E}_{\text{field}}(\boldsymbol{r},t)$ is dependent solely on the local and instantaneous values of the fields, $\boldsymbol{E}(\boldsymbol{r},t)$ and $\boldsymbol{H}(\boldsymbol{r},t)$, and the sources, $\boldsymbol{J}_{\text{free}}(\boldsymbol{r},t)$, $\boldsymbol{P}(\boldsymbol{r},t)$ and $\boldsymbol{M}(\boldsymbol{r},t)$. Since Eq.(11) specifies only the time-rate-of-change of $\mathcal{E}_{\text{field}}(\boldsymbol{r},t)$, the energy density itself may depend on the history of the local fields. Also, since $\boldsymbol{P}(\boldsymbol{r},t)$ and $\boldsymbol{M}(\boldsymbol{r},t)$ could, in principle, depend on the $E$- and $H$-fields elsewhere in time and space, the energy density's dependence on the fields may or may not be local.

Note that, as far as energy density is concerned, the bound electric current $\boldsymbol{J}_{\text{e\_bound}} = \partial \boldsymbol{P}(\boldsymbol{r},t)/\partial t$ behaves similarly to $\boldsymbol{J}_{\text{free}}$ in response to $\boldsymbol{E}(\boldsymbol{r},t)$, whereas $\boldsymbol{J}_{\text{e\_mag}} = \mu_o^{-1} \nabla \times \boldsymbol{M}$ does not enter the above expression at all. What shows up in the energy density expression is $\boldsymbol{J}_{\text{m\_bound}} = \partial \boldsymbol{M}(\boldsymbol{r},t)/\partial t$; however, this "magnetic current" interacts with the $H$-field rather than with the $E$-field. In any event, so long as Eq.(11) is accepted as a *postulate* of the classical theory of electromagnetism, there is no need to speculate about the meaning of its various terms. The sole justification for the energy postulate of Eq.(11) is that its predictions and consequences remain consistent with the law of conservation of energy as well as with experimental findings.

Depending on the sign of $\boldsymbol{E} \cdot \boldsymbol{J}_{\text{free}}$ in Eq.(11), the free current's contribution to $\partial \mathcal{E}_{\text{field}}/\partial t$ could be positive or negative. In other words, $\boldsymbol{J}_{\text{free}}$ absorbs energy from the field when the local $E$-field's projection on $\boldsymbol{J}_{\text{free}}$ is positive, whereas the energy flows in the reverse direction – from the current to the field – when the projection of $\boldsymbol{E}$ on $\boldsymbol{J}_{\text{free}}$ is negative. The local, instantaneous energy density stored in $E$- and $H$-fields is seen from Eq.(11) to be $\frac{1}{2}\varepsilon_0|\boldsymbol{E}|^2 + \frac{1}{2}\mu_0|\boldsymbol{H}|^2$; this energy may rise or fall with time, but is never converted *directly* to heat, mechanical work, etc. (We mention in passing that, to our best knowledge, the vacuum energy density of time-dependent $E$- and $H$-fields, namely, $\mathcal{E}_{\text{field}}(\boldsymbol{r},t) = \frac{1}{2}\varepsilon_0|\boldsymbol{E}(\boldsymbol{r},t)|^2 + \frac{1}{2}\mu_0|\boldsymbol{H}(\boldsymbol{r},t)|^2$, has never been derived from first principles; as such, even in standard treatments of the classical theory, this part of Eq.(11) must be taken as an independent postulate rather than a consequence of Maxwell's equations.)

Another contribution to the right-hand side of Eq.(11) comes from $\boldsymbol{E} \cdot \partial \boldsymbol{P}/\partial t$; when positive, this term expresses the rate at which energy is stored in the polarization $\boldsymbol{P}$; when negative, it represents the rate of return of energy from $\boldsymbol{P}$ to the field. In general, $\boldsymbol{P}(\boldsymbol{r},t)$ can serve either as a lossless or a lossy reservoir of energy, or even as a source of energy (e.g., in gain media). The remaining term on the right-hand side of Eq.(11), $\boldsymbol{H} \cdot \partial \boldsymbol{M}/\partial t$, behaves similarly to $\boldsymbol{E} \cdot \partial \boldsymbol{P}/\partial t$, with the obvious difference that the exchange of energy between the fields and the magnetization $\boldsymbol{M}$ is mediated by the $H$-field rather than the $E$-field.

One may now proceed to dot-multiply Maxwell's second and third equations, Eqs.(1b) and (1c), with $\boldsymbol{E}(\boldsymbol{r},t)$ and $-\boldsymbol{H}(\boldsymbol{r},t)$, respectively, then add the resulting equations and invoke Eq.(11) to arrive at Poynting's theorem, namely,

$$\nabla \cdot \boldsymbol{S}(\boldsymbol{r},t) = \nabla \cdot [\boldsymbol{E}(\boldsymbol{r},t) \times \boldsymbol{H}(\boldsymbol{r},t)] = -\partial \mathcal{E}_{\text{field}}(\boldsymbol{r},t)/\partial t. \tag{12}$$

Although Eq.(12) does not identify a unique Poynting vector – in the sense that any divergence-free vector field could be added to $\boldsymbol{E}(\boldsymbol{r},t) \times \boldsymbol{H}(\boldsymbol{r},t)$ without modifying the content of the equation – we believe, along with Richard Feynman [2], that the simplest choice, namely, $\boldsymbol{S}(\boldsymbol{r},t) = \boldsymbol{E}(\boldsymbol{r},t) \times \boldsymbol{H}(\boldsymbol{r},t)$, yields the most physically meaningful expression for the rate of flow of energy per unit area per unit time. In fact, to avoid such ambiguities, it is perhaps preferable to replace the energy postulate of Eq.(11) with the following, slightly more general, postulate:

*The rate of flow of energy per unit area per unit time is the Poynting vector $\boldsymbol{S}(\boldsymbol{r},t) = \boldsymbol{E}(\boldsymbol{r},t) \times \boldsymbol{H}(\boldsymbol{r},t)$.*

A direct consequence of this postulate (in conjunction with Maxwell's macroscopic equations) will then be the expression of the time-rate-of-change of energy density given by Eq.(11).

**8. Linear and angular momenta of the electromagnetic field**. Another tenet of the classical theory that needs specific enunciation is the expression of electromagnetic momentum density $\boldsymbol{p}_{EM}(\boldsymbol{r}, t)$ in terms of the Poynting vector $\boldsymbol{S}(\boldsymbol{r}, t)$. There exist several thought experiments, collectively referred to as "Einstein Box" experiments, that relate the electromagnetic field's momentum density to the Poynting vector [2,16]. These arguments are not sufficiently general and, moreover, involve certain assumptions that lie outside the domain of classical theory; as such, it is preferable to treat electromagnetic momentum via a postulate that applies not only in vacuum but also in the presence of $\rho_{\text{free}}$ and $\boldsymbol{J}_{\text{free}}$, and in material media that possess electric and/or magnetic polarization, $\boldsymbol{P}$ and $\boldsymbol{M}$. The general expression for the density of electromagnetic momentum (also known as the Abraham momentum [5-7]) is

$$\boldsymbol{p}_{EM}(\boldsymbol{r}, t) = \boldsymbol{S}(\boldsymbol{r}, t)/c^2. \tag{13a}$$

The corresponding formula for the electromagnetic angular momentum density $\boldsymbol{L}_{EM}(\boldsymbol{r}, t)$ is

$$\boldsymbol{L}_{EM}(\boldsymbol{r}, t) = \boldsymbol{r} \times \boldsymbol{p}_{EM}(\boldsymbol{r}, t) = \boldsymbol{r} \times \boldsymbol{S}(\boldsymbol{r}, t)/c^2. \tag{13b}$$

The above expressions apply not only to propagating fields in vacuum, but also to static fields and evanescent fields, as well as fields within media that contain charge, current, polarization, and magnetization, whether or not these media are homogeneous, transparent, isotropic, linear, etc. Moreover, the expression of the angular momentum density in Eq.(13b) applies to both spin and orbital angular momenta; in other words, the formula is valid irrespective of whether the angular momentum arises from the polarization state of the field, from its phase and amplitude profile (e.g., field vorticity), or from a combination of the two.

When a pulse of light propagates in the free space, its electromagnetic momentum (linear or angular) remains the same at all times. However, once the pulse encounters a material medium and begins to scatter from or enter into that medium, the total electromagnetic momentum of the system (linear or angular) begins to change with time. The time-rate-of-change of the total linear (angular) electromagnetic momentum is exactly equal and opposite to the total force (torque) exerted by the light on the material medium. This is a general statement of the law of conservation of linear (angular) momentum. Verifying the above statement, however, requires a knowledge of the force (torque) exerted by the electromagnetic field on material media; this is the subject of the following section.

**9. Force and torque exerted by the electromagnetic field on material media**. To complete the foundational postulates of the classical theory of electromagnetism, it remains to express the force and torque densities exerted by the electromagnetic field on material media. This we do in the case of media defined by their $\rho_{\text{free}}(\boldsymbol{r}, t)$, $\boldsymbol{J}_{\text{free}}(\boldsymbol{r}, t)$, $\boldsymbol{P}(\boldsymbol{r}, t)$ and $\boldsymbol{M}(\boldsymbol{r}, t)$. The generalized expressions of the Lorentz force and torque densities are

$$\boldsymbol{F}_1(\boldsymbol{r},t) = \rho_{\text{free}}\boldsymbol{E} + \boldsymbol{J}_{\text{free}} \times \mu_o\boldsymbol{H} + (\boldsymbol{P} \cdot \nabla)\boldsymbol{E} + (\boldsymbol{M} \cdot \nabla)\boldsymbol{H} + (\partial \boldsymbol{P}/\partial t) \times \mu_o\boldsymbol{H} - (\partial \boldsymbol{M}/\partial t) \times \varepsilon_o\boldsymbol{E}. \tag{14a}$$

$$\boldsymbol{T}_1(\boldsymbol{r},t) = \boldsymbol{r} \times \boldsymbol{F}_1(\boldsymbol{r},t) + \boldsymbol{P}(\boldsymbol{r},t) \times \boldsymbol{E}(\boldsymbol{r},t) + \boldsymbol{M}(\boldsymbol{r},t) \times \boldsymbol{H}(\boldsymbol{r},t). \tag{14b}$$

Using simple examples that are amenable to exact analysis, we have shown in previous publications [16-19] that Eqs.(14) lead to a precise balance of linear and angular momenta when all relevant forces, especially those at the boundaries, are properly taken into account. Similar or even identical expressions for force and torque densities have been derived by others [3,21-24]. Our focus, however, has been the generalization of the Lorentz law in a way that is consistent with Maxwell's equations, with the principles of special relativity, and with the conservation laws, *without* regard for the underlying physical mechanisms that give rise to $\boldsymbol{P}$ and $\boldsymbol{M}$.

In evaluating the force and torque exerted by the electromagnetic field on ponderable media, care must be taken that, in every instance, the relevant equations are solved self-consistently. For example, any motion imparted to the medium in consequence of the exertion of electromagnetic force and torque which would result in a change of the spatio-temporal dependence of $\rho_{\text{free}}$, $\boldsymbol{J}_{\text{free}}$, $\boldsymbol{P}$ and $\boldsymbol{M}$ must be automatically incorporated into the solution of Maxwell's equations, solutions that relate the $\boldsymbol{E}, \boldsymbol{D}, \boldsymbol{H}, \boldsymbol{B}$ fields to their sources $\rho_{\text{free}}, \boldsymbol{J}_{\text{free}}, \boldsymbol{P}, \boldsymbol{M}$. Alternatively, if the electromagnetic fields are computed by assuming the sources $\rho_{\text{free}}(\boldsymbol{r}, t)$, $\boldsymbol{J}_{\text{free}}(\boldsymbol{r}, t)$, $\boldsymbol{P}(\boldsymbol{r}, t)$ and $\boldsymbol{M}(\boldsymbol{r}, t)$, then the resulting force and torque on these sources, computed in accordance with Eqs.(14), cannot be allowed to further modify the sources; in particular, the exerted electromagnetic force and torque should not result in *additional* material motion, acoustic wave generation and propagation, etc., in a way that would modify the assumed strengths of the sources or their spatio-temporal dependences. This is not to say that mechanical motion and acoustic wave propagation should be ignored; rather it is stating the obvious that such motion must be treated self-consistently.

**10. Alternative expression for force and torque densities.** There exists an alternative formulation of the generalized Lorentz law, where bound electric and magnetic charge densities $\rho_{\text{e\_bound}} = -\nabla \cdot \boldsymbol{P}$ and $\rho_{\text{m\_bound}} = -\nabla \cdot \boldsymbol{M}$ directly experience the force of the $\boldsymbol{E}$ and $\boldsymbol{H}$ fields. The alternative formulas for the force and torque densities are

$$\boldsymbol{F}_2(\boldsymbol{r},t) = (\rho_{\text{free}} - \nabla \cdot \boldsymbol{P})\boldsymbol{E} + (\boldsymbol{J}_{\text{free}} + \partial \boldsymbol{P}/\partial t) \times \mu_o \boldsymbol{H} - (\nabla \cdot \boldsymbol{M})\boldsymbol{H} - (\partial \boldsymbol{M}/\partial t) \times \varepsilon_o \boldsymbol{E}. \tag{15a}$$

$$\boldsymbol{T}_2(\boldsymbol{r},t) = \boldsymbol{r} \times \boldsymbol{F}_2(\boldsymbol{r},t). \tag{15b}$$

The equivalence of total force (and total torque) for the two formulations in Eqs. (14) and (15) is implicit in the analysis of Hansen and Yaghjian [3], but was proven explicitly (and independently) by Barnett and Loudon [25, 26]. Subsequently, we extended the proof to cover the case of objects immersed in a liquid [27, 28]. As far as the total force (or total torque) exerted on a given volume of material is concerned, Eqs. (14) and (15) can be shown to yield identical results provided that forces at the boundaries are properly treated in each case in accordance with the corresponding force equation. The force (torque) distribution throughout the volume, of course, will depend on which formulation is used, but when integrated over the volume of interest, the two distributions always yield identical values for total force (and total torque).

Although, mathematically speaking, both formulations of the generalized Lorentz law given in Eqs. (14) and (15) are acceptable, when it comes to real-world physical problems only one formulation should apply in any given situation. It is conceivable of course that, depending on the physical mechanisms that underlie $\boldsymbol{P}$ and $\boldsymbol{M}$, some material media will exhibit the dipolar behavior embodied in Eqs. (14), while others will behave in accordance with the bound-charge model of Eqs. (15). Either way, the best approach to deciding between the two formulations appears to be conducting experiments that would unambiguously determine the distribution of force and torque throughout the volume of a material body exposed to electromagnetic radiation.

Looking at the problem from this experimental perspective, one can argue that Eqs. (14) are superior to Eqs. (15), as the former already incorporate the latter. To appreciate this argument, note that ($\rho_{\text{free}}$, $\boldsymbol{J}_{\text{free}}$) is one source of electromagnetic fields, while $\boldsymbol{P}(\boldsymbol{r}, t)$ is another, and $\boldsymbol{M}(\boldsymbol{r}, t)$ is yet a third source. In real-world physical systems, one generally associates the behavior of material media with one or more of these sources. Now, technically speaking, in solids and liquids none of the charges are free, as they are all bound with the lattice. However, those electrons that are tightly bound with their host nuclei can be said to form electric dipoles; the force exerted on these dipoles by the $E$-field is $(\boldsymbol{P} \cdot \nabla)\boldsymbol{E}$, as in Eq. (14a). On the other hand, electrons that are more or less free to roam around the lattice (e.g., conduction electrons) may be said to act as free electrons; for these the force exerted by the $E$-field is $-(\nabla \cdot \boldsymbol{P})\boldsymbol{E}$, as in Eq. (15a). However, this last term would be readily present in Eq. (14a) if, by definition, $\rho_{\text{free}}$ were to contain $-\nabla \cdot \boldsymbol{P}$. (It may seem strange to think of a conduction electron as belonging to a dipole; however, it is well-known that the Lorentz oscillator model treats conduction electrons very much like bound electrons, except that the spring constant connecting a conduction electron to its host atom is set to zero. This, of course, makes sense for all frequencies except $\omega = 0$, where under a constant $E$-field the electron would drift away from its host atom.)

One may imagine that, if there are electrons in the lattice which are neither tightly-bound nor entirely free, they will spend a fraction of their time acting like free electrons, while the remainder of their time is spent in the bound state. It will then be possible to model these electrons as a mixture of ($\rho_{\text{free}}$, $\boldsymbol{J}_{\text{free}}$) and $\boldsymbol{P}(\boldsymbol{r}, t)$. The task of modeling this type of material thus involves a determination as to what fraction of the charges act as free and what fraction as bound, in order to assign appropriate numerical values to ($\rho_{\text{free}}$, $\boldsymbol{J}_{\text{free}}$) and $\boldsymbol{P}(\boldsymbol{r}, t)$. The beauty of the force and torque expressions in Eqs. (14) is that they allow for both types of behavior, whereas Eqs. (15) do not distinguish the essential dipolar nature of $\boldsymbol{P}(\boldsymbol{r}, t)$ from the free (or loosely-bound) character of ($\rho_{\text{free}}$, $\boldsymbol{J}_{\text{free}}$).

The same sort of argument can be made for the force and torque exerted on $\boldsymbol{M}(\boldsymbol{r}, t)$ in Eqs. (14) and (15). Of course, since magnetic monopoles have not been observed in Nature, we do not have anything equivalent to ($\rho_{\text{free}}$, $\boldsymbol{J}_{\text{free}}$) in this case. However, there is no a priori reason to believe that some fraction of the magnetic "dipoles" in real world will not behave like a pair of loosely-bound monopoles, in which case the force exerted upon them by an applied $H$-field should be modeled as $-(\nabla \cdot \boldsymbol{M})\boldsymbol{H}$, whereas a tightly-bound magnetic dipole would respond to an applied $H$-field in accordance with the $(\boldsymbol{M} \cdot \nabla)\boldsymbol{H}$ formulation. Once an experimental determination has been made as to what fraction of $\boldsymbol{M}(\boldsymbol{r}, t)$, if any, behaves as a free magnetic charge density, the corresponding force density, $\rho_{\text{m-free}}\boldsymbol{H} - \boldsymbol{J}_{\text{m-free}} \times \varepsilon_o \boldsymbol{E}$, should be added to Eq. (14a) to account for this behavior.

**11. Does the field momentum-energy constitute a 4-vector?** In special relativity the momentum and energy of a particle (or system of particles) form a 4-vector ($\boldsymbol{p}$, $\mathcal{E}/c$). This property is also shared by isolated pulses of electromagnetic radiation traveling in the free space. However, when the electromagnetic field is "attached" to its source(s), it is impossible to separate the momentum-energy of the field from that of the source in order to define a stand-alone momentum-energy 4-vector for the field. The following examples demonstrate this point.

Consider the $L \times L \times d$ capacitor depicted in Fig. 1(a), and assume $L \gg d$ so that the edge effects can be ignored. In the $xyz$ frame, where the capacitor is stationary, the uniform charge densities on the two plates are $\pm \sigma_o$, the $E$-field is $(\sigma_o/\varepsilon_o)\hat{y}$, and the $H$-field is zero. The total field energy is thus $\tfrac{1}{2}(\sigma_o^2/\varepsilon_o)L^2 d$, and, with the absence of the $H$-field implying the absence of the Poynting vector $S$, the field momentum is zero. The scalar and vector potentials in the $xyz$ frame in the region between the plates are $\psi(y) = -(\sigma_o/\varepsilon_o)y$ and $A(r,t) = 0$. In the $x'y'z'$ frame, where the capacitor travels with velocity $V$ along the $x$-axis, the potentials are $\psi'(y') = -\gamma(\sigma_o/\varepsilon_o)y'$ and $A'(y') = -\gamma(V/c^2)(\sigma_o/\varepsilon_o)y'\hat{x}$. The fields are thus $E' = \gamma(\sigma_o/\varepsilon_o)\hat{y}$ and $H' = \gamma\sigma_o V\hat{z}$, and the total field energy and momentum are $\tfrac{1}{2}\gamma(\sigma_o^2/\varepsilon_o)(1+V^2/c^2)L^2 d$ and $\gamma(\sigma_o^2/\varepsilon_o)(V/c^2)L^2 d\hat{x}$, respectively. (In the latter expressions, the FitzGerlad-Lorentz contraction of the moving capacitor along the $x$-axis has been taken into account.) Clearly the field energy and momentum in the two frames are *not* related through a Lorentz transformation.

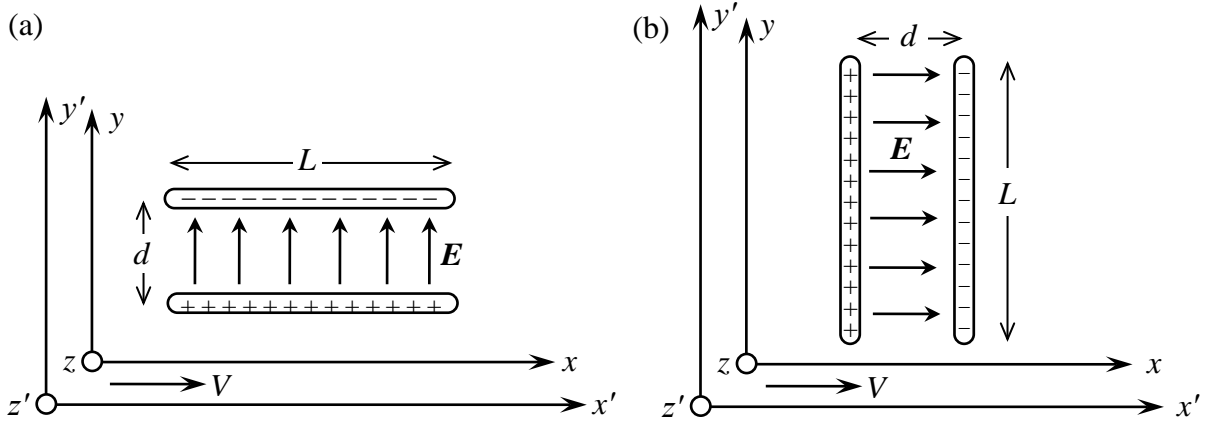

**Figure 1**. A capacitor consisting of two $L \times L$ plates separated by a distance $d$ in free space is uniformly charged with a surface charge density of $\pm \sigma_o$. The capacitor is stationary in the $xyz$ frame, and moves with velocity $V$ along the $x$-axis in the $x'y'z'$ frame. The plates are parallel to the $xz$-plane in (a) and perpendicular to it in (b).

The situation in Fig. 1(b) is similar to that depicted in Fig. 1(a), except that the capacitor is now rotated around the $z$-axis by 90°. In the $xyz$ frame, the fields as well as their energy and momentum are the same as before; the scalar potential in the region between the plates, however, has become $\psi(x) = -(\sigma_o/\varepsilon_o)x$. In the $x'y'z'$ frame the potentials are $\psi'(x',t') = -\gamma^2(\sigma_o/\varepsilon_o)(x'-Vt')$ and $A'(x',t') = -\gamma^2(V/c^2)(\sigma_o/\varepsilon_o)(x'-Vt')\hat{x}$. Thus the fields in $x'y'z'$ are $E' = (\sigma_o/\varepsilon_o)\hat{x}$ and $H' = 0$; the field's total energy is $\tfrac{1}{2}\gamma(\sigma_o^2/\varepsilon_o)(1-V^2/c^2)L^2 d$ while its momentum is zero. (In the energy expression, the FitzGerlad-Lorentz contraction of the gap between the capacitor plates has been taken into account.) Once again, in going from one inertial frame to another, the field's momentum-energy is seen to behave in a way that is *not* expected from a 4-vector.

In the above examples, the departure of the field's momentum-energy from 4-vector behavior must be related to the hitherto ignored momentum-energy contributions of the (charged) capacitor plates. If, in the stationary ($xyz$) frame, the energy of the material part of the system is denoted by $\mathcal{E}_o L^2 d$, then, in the moving ($x'y'z'$) frame, the *total* energy and momentum of the system become $\gamma(\mathcal{E}_o + \tfrac{1}{2}\sigma_o^2/\varepsilon_o)L^2 d$ and $\gamma(\mathcal{E}_o + \tfrac{1}{2}\sigma_o^2/\varepsilon_o)(V/c^2)L^2 d\hat{x}$, respectively. Comparing these with the field energy and momentum in the preceding examples, we find that the energy and momentum of the material part of the moving capacitor must be given by $\gamma[\mathcal{E}_o \pm \tfrac{1}{2}(\sigma_o^2/\varepsilon_o)(V/c)^2]L^2 d$ and $\gamma[\mathcal{E}_o \pm \tfrac{1}{2}(\sigma_o^2/\varepsilon_o)](V/c^2)L^2 d\hat{x}$, with the minus sign applying to the case depicted in Fig. 1(a), and the plus sign to that in Fig. 1(b). Clearly the momentum-energy of neither the "material part" nor the "field part" of the capacitor exhibits 4-vector behavior; only the sum total of these parts behaves in a way that is consistent with special relativity.

**12. Numerical simulations**. We present a set of numerical results that serves to illustrate some of the statements made regarding electromagnetic force and momentum in the preceding sections. The sequence of Finite Difference Time Domain (FDTD) simulation results depicted in Figs. 2(a-f) shows the propagation of a short pulse of light in the free space, its interaction with a transparent prism followed by interaction with a partially absorbing reflector, and the subsequent passage of the reflected pulse through the glass prism. In this two-dimensional FDTD simulation, the incident light pulse is linearly polarized, with its $E$-field in the $yz$-plane. The various plots represent the magnitude of the component $E_y$ of the electric field parallel to the $y$-axis.

Figure 2(a) shows a short pulse of light propagating in the free space along the negative $z$-axis. The pulse first arrives at a dielectric prism of refractive index $n = 1.5$ whose outline appears in dashed white lines. A small fraction of the pulse's energy bounces off the entrance facet, while the rest enters the prism. The pulse is then reflected from the left-hand side facet, and exits the prism on the right-hand side. Initially, some of the pulse's energy goes into an evanescent field, but, eventually, this field separates itself from the prism and propagates away. The object outlined in dashed black lines in Fig. 2 is a homogeneous medium that partially reflects and partially absorbs the incident light. The Debye dispersion model used for this partial mirror is $\varepsilon_{\text{mirror}}(\omega) = \varepsilon_\infty + \Delta\varepsilon/(1+i\omega\tau)$, with $\varepsilon_\infty = 2.0$, $\Delta\varepsilon = 20.0$, $\tau = 4.8437$ fs. While a small fraction of the pulse's energy is absorbed within the mirror, a major portion of the light bounces off the mirror's reflecting surface, goes back through the dielectric prism, and returns to the free space in the end. Each frame of Fig. 2 depicts a $12 \times 12\,\mu\text{m}^2$ region within the $yz$-plane. Different frames represent different instants of time, starting at $t = 0$, when the pulse begins its journey in the region immediately above the prism, and ending at $t = 27.5$ fs, when the pulse is substantially broken up via scattering, reflection, absorption, and diffraction processes. Throughout this propagation-and-scattering process, we kept track of the total electromagnetic momentum and total force exerted by the light pulse on the material objects within the system.

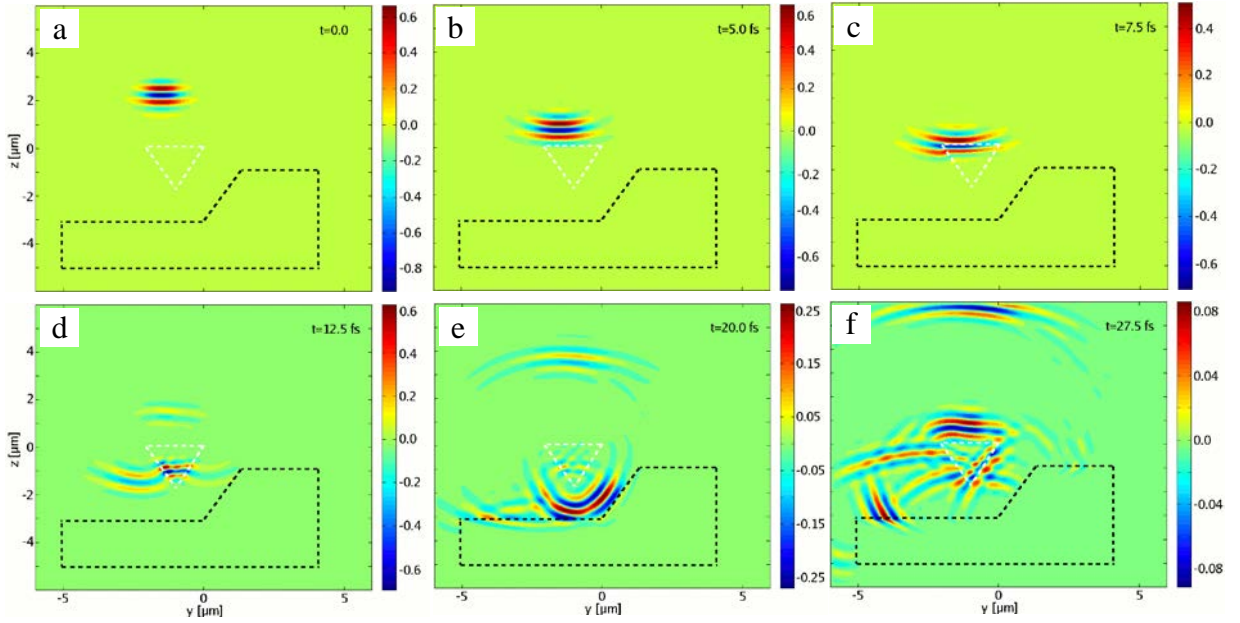

**Figure 2**. A short pulse of light propagating in free space along the negative $z$-axis arrives at a dielectric prism whose outline is shown in dashed white lines. The object outlined in dashed black lines is a homogeneous medium that partially reflects and partially absorbs the incident light. Each frame depicts a $12 \times 12\,\mu\text{m}^2$ region in the $yz$-plane. Different frames represent different instants of time, starting at $t = 0$, when the pulse begins its journey in the region immediately above the prism, and ending at $t = 27.5$ fs, when the pulse is substantially broken up via scattering, reflection, absorption, and diffraction processes. In this two-dimensional FDTD simulation, the incident light pulse is linearly polarized, with its $E$-field in the $yz$-plane. The various plots represent the magnitude of the component $E_y$ of the $E$-field parallel to the $y$-axis.

Figure 3 shows the time evolution of total electromagnetic momentum $\boldsymbol{p}_{\text{EM}}(t) = (1/c^2)\iint \boldsymbol{S}(y,z,t)\,\text{d}y\,\text{d}z$ and total force $\boldsymbol{F}(t) = \iint \boldsymbol{F}(y,z,t)\,\text{d}y\,\text{d}z$ exerted by the light pulse on the material media in the system of Fig. 2 during the first 25 fs of the process. The red (–·–) and blue (–··–) curves show the evolution of $p_y$ and $p_z$, respectively, while the black (——) and green (– – –) curves are the computed force components, $F_y$ and $F_z$. The solid circles superimposed on the $F_y(t)$ and $F_z(t)$ plots represent time-derivatives of $-p_y(t)$ and $-p_z(t)$, thus confirming the universal relation $\boldsymbol{F}(t) = -\text{d}\boldsymbol{p}_{\text{EM}}(t)/\text{d}t$.

It is clear that the time-rate-of-change of *total* linear momentum is exactly equal (and opposite) to the *total* force exerted by the light pulse on the material media. This general equivalence holds at all instants of time and is apparently valid even though some of the electromagnetic momentum happens to reside in the free space, in the form of propagating as well as evanescent waves, while some fraction of the momentum resides inside transparent as well as absorbing and dispersive media.

**13. Concluding remarks**. In this paper we argued that the macroscopic equations of Maxwell are a consistent and mathematically precise set of equations that can be used to analyze general problems in classical electrodynamics *without* the need for specific physical models of polarization density $\boldsymbol{P}$ and magnetization density ***M***. The standard Maxwell equations, however, must be augmented with two postulates regarding the energy and momentum of the field, and also with a generalized form of the Lorentz force law, in order to provide a complete and consistent set of equations that comply not only with the requirements of the special theory of relativity, but also with the laws of conservation of energy and momentum.

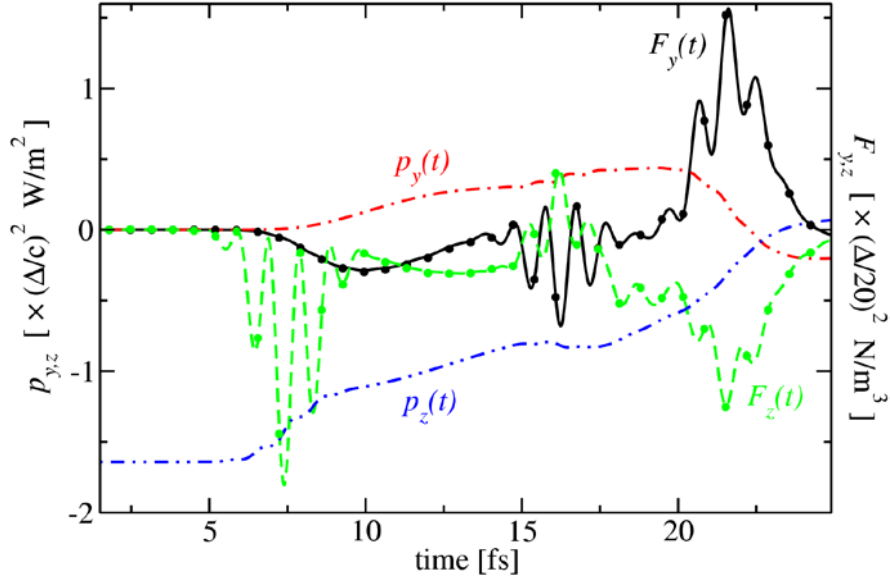

**Figure 3**. Evolution of total electromagnetic momentum ***p***$_{EM}$(*t*) and total force ***F***(*t*) exerted by the light pulse on the material media during the process depicted in Fig. 2. The red (–·–) and blue (–··–) curves show the evolution of $p_y$ and $p_z$; black (——) and green (– – –) curves show the computed force components $F_y$ and $F_z$. The solid circles superposed on $F_y(t)$ and $F_z(t)$ are time-derivatives of $-p_y(t)$ and $-p_z(t)$.

With the aid of the momentum postulate, in particular, we have argued that the long-standing Abraham-Minkowski controversy surrounding the momentum of the electromagnetic field inside material media can be resolved. In general, interactions between the field and its material environment result in a change of the total field momentum versus time. The field may be distributed among different parts of the host medium, with perhaps some fraction of it residing in the surrounding free space, in the form of propagating and/or evanescent waves. The momentum postulate fixes the momentum density at ***S***(***r***, *t*)/*c*$^2$ at each and every point in the system where the field exists. The time-rate-of-change of the total field momentum then yields the total force experienced by the host medium. In the past, the force of the electromagnetic field on the host medium has sometimes been wholly or partially attributed to a "mechanical" momentum accompanying the field momentum [7,8,16]. Such distinctions, however, are no longer necessary in light of the arguments presented in this paper. Our conclusions with regard to momentum may be summarized as follows:

i) Abraham momentum is the sole electromagnetic momentum in any system of materials distributed throughout the free space.

ii) Force and torque densities may be directly and unambiguously computed from the generalized Lorentz law.

iii) The total instantaneous force (torque) is precisely equal and opposite to the time-rate-of-change of the total electromagnetic linear (angular) momentum of the system.

The last assertion is simply a statement of the laws of conservation of linear and angular momenta.

Finally, although we have been able to demonstrate, either analytically or numerically, the conservation of energy and momentum in diverse situations involving linear, dispersive, and isotropic as well as birefringent media [8-19], the most general proof of the statements made in the preceding sections (when arbitrary polarization and/or magnetization functions are involved) has not yet been attempted. As likely as it seems, based on our extensive calculations and simulations, that the postulates will hold under general circumstances, it will be desirable to verify the conjectures of this paper in new cases that will turn out to be amenable to analytical or numerical investigation. A good example of such cases is the transmission of light through silica nano-fibers, as reported by She *et al* in a recent publication [29]. Although we are encouraged by the authors' claim that their experimental findings confirm the hypothesis that photons within the nano-fiber have the Abraham momentum, we believe it is necessary to carry out precise numerical calculations in order to properly account for the electromagnetic momentum both inside and outside the nano-fiber, as well as the Lorentz force exerted by the light pulse in its *entire* path through the nano-fiber waveguide.

**Acknowledgements.** This work has been supported by the Air Force Office of Scientific Research (AFOSR) under contract number FA 9550−04−1−0213. The authors are grateful to Ewan M. Wright, Miroslav Kolesik, and Arthur D. Yaghjian for many helpful discussions.